\pdfoutput=1
\documentclass[11pt,a4paper]{article}
\usepackage{jheppub}
\usepackage[caption=false]{subfig}
\usepackage{graphicx}
\usepackage{hyperref}
\usepackage{multirow}
\usepackage{color}

\def\p{\partial}

\def\=:{=\hspace{-.7em}\raisebox{1.1ex}{.}\hspace{.1em}\raisebox{-0.2ex}{.}}
\newcommand{\tr}{{\rm tr}\,}

\newcommand {\beq}{\begin{eqnarray}}
\newcommand {\eeq}{\end{eqnarray}}
\newcommand {\non}{\nonumber\\}

\begin{document}

\title{Black hole Skyrmion in a generalized Skyrme model}

\author{Sven Bjarke Gudnason,${}^1$}
\author{Muneto Nitta${}^2$ and}
\author{Nobuyuki Sawado${}^3$}
\affiliation{${}^1$Institute of Modern Physics, Chinese Academy of
  Sciences, Lanzhou 730000, China}
\affiliation{${}^2$Department of Physics, and Research and Education
  Center for Natural Sciences, Keio University, Hiyoshi 4-1-1,
  Yokohama, Kanagawa 223-8521, Japan}
\affiliation{${}^3$Department of Physics, Tokyo University of Science,
  Noda, Chiba 278-8510, Japan}
\emailAdd{bjarke(at)impcas.ac.cn}
\emailAdd{nitta(at)phys-h.keio.ac.jp}
\emailAdd{sawado(at)ph.noda.tus.ac.jp}

\abstract{
We study a Skyrme-like model with the Skyrme term and a sixth-order
derivative term as higher-order terms, coupled to gravity and we
construct Schwarzschild black hole Skyrme hair.
We find, surprisingly, that the sixth-order derivative term alone
cannot stabilize the black hole hair solutions; the Skyrme term with a
large enough coefficient is a necessity.
}


\maketitle

\section{Introduction}

Black holes are generally believed to be characterized by two
properties at asymptotically far distances, namely their masses and
their global charges.
This is known is the weak no-hair conjecture.
The first stable counter example to this no-hair conjecture was made
in the framework of the Skyrme model, i.e.~a black hole with Skyrme
hair
\cite{Luckock:1986tr,Glendenning:1988qy,Droz:1991cx,Heusler:1991xx,Heusler:1992av,Bizon:1992gb,Volkov:1998cc}
(for a review, see Ref.~\cite{Shiiki:2005pb}).
The Skyrme model is a scalar field theory based on the chiral
Lagrangian with the addition of a quartic derivative term, which is
like a curvature term on the internal (target) space of the model
\cite{Skyrme:1962vh,Skyrme:1961vq}. 
In flat space the Skyrmion is a map from space ($\mathbb{R}^3$) to an
SU(2) target space, which is characterized by $\pi_3(S^3)$, giving
rise to topological solitons; i.e.~the Skyrmions.
As implied by the fact that they are topological, their charges --
called baryon charges -- are integers in flat space.
The Skyrmion with a black hole is interpreted as a black hole with
scalar hair and the asymptotic behavior of the Skyrmion is very
similar to that in flat space.
Near the black hole the Skyrmion is deformed, nevertheless. 
Now when the black hole is formed with the Skyrmion surrounding it,
the Skyrmion loses a fraction of its charge; this happens due to the
fact that the profile function of the Skyrmion, which normally
``winds'' $\pi$ to complete the 3-cycle, only ``winds''
$\pi-\epsilon$.  
When the black hole horizon -- and hence its mass -- becomes larger
than a certain critical value then the Skyrmion ceases to exist and
the Skyrme hair becomes unstable. 

Another twist to the Skyrmion solution in flat space is that when it
is the hair of a black hole, two branches of solutions (fixed points
of the action) open up \cite{Luckock:1986tr}; one of these two
branches of solutions contains, however, unstable Skyrmion solutions.
The two branches bifurcate at the above mentioned critical mass or
horizon radius, beyond which no stable solution exists.
If we pick a point on the stable branch and take the limit of the
black hole mass going to zero, then the solutions converge smoothly
to that of the flat space.
If we now pick a point on the unstable branch, the answer depends on
whether the gravitational coupling is turned on or not; if it is
turned on -- which is tantamount to the gravitational backreaction
being taken into account -- then the conclusion remains the same; the
solution converges to that of flat space. If the gravitational
coupling is turned off, however, then the solution becomes
discontinuous and ceases to exist -- the limit is hence not well
defined. 

Apart from the seminal result of Luckock
et.al.~(Ref.~\cite{Luckock:1986tr}), and the papers that followed;
other variants of the Skyrme black hole hair system have been studied
in the literature.  
The most natural generalization is to turn on a nonvanishing
cosmological constant; in Refs.~\cite{Shiiki:2005aq,Shiiki:2005xn} and 
\cite{Brihaye:2005an} the black hole Skyrme hair was ported to anti-de
Sitter and de Sitter spacetimes, respectively.
The late-time evolution of the radiation emitted from the black hole
with Skyrme hair was studied in
Refs.~\cite{Zajac:2009am,Zajac:2010mu}.
Gravitating sphalerons in the Einstein-Skyrme model have been
constructed in Ref.~\cite{Shnir:2015aba}.
Quantization of collective coordinates in the Skyrmion black hole was
carried out \cite{Shiiki:2004jj}. 
Another natural generalization of the black hole Skyrmion system, is
to consider the Skyrmion surrounding the black hole to have a higher 
charge (winding); a particular class of axially symmetric solutions
has been found in Refs.~\cite{Sawado:2003at,Sawado:2004yq} and
quantization of collective coordinates was considered in such
systems as well \cite{Sato:2006xy}.
Recently, it has been contemplated that black holes do not necessarily
violate the baryon number when the possibility of black hole
Skyrme hair is taken into consideration \cite{Dvali:2016mur}. 

An interesting question is: what is the foundation of the stabilizing 
mechanism of the black hole hair? If we turn off the Skyrme term, the
scalar hair is not stable.
In light of recent developments in the Skyrme model, which was
motivated by a completely different effect -- namely the large binding 
energy of the multi-Skyrmion, being too large for the Skyrmions to be 
interpreted as nuclei -- a sixth-order derivative term has been
introduced \cite{Adam:2010fg,Adam:2010ds} and this model has been
dubbed the BPS-Skyrme model (neutron stars have been studied in the
framework of the BPS-Skyrme model \cite{Adam:2014dqa,Adam:2015lpa} and
we have recently found gravitating analytic and numerical Skyrmion
solutions in the BPS-Skyrme model \cite{Gudnason:2015dca}). 
For the story of the binding energy, the sextic term has the
interesting property that a saturable BPS bound exists in the subset
of the model containing only said sextic term as well as a potential
term.
Using Derrick's theorem \cite{Derrick:1964ww}, any higher-order
derivative term can stabilize the Skyrmion solution in flat space, by
balancing the pressure with respect to that of the kinetic (Dirichlet)
term and/or the potential. 
As for the hair of the black hole, however, it is far less trivial
which kind of terms can stabilize the black hole hair. 
One may naively think that we can substitute the Skyrme term with the
sixth-order derivative term and retain a similar black hole hair
solution.
Our findings, however, suggest otherwise.
Although we can add the sextic term to the model and have stable black 
hole hair; the Skyrme term with a positive coefficient is a
necessity. 
This is the main result of our paper. 

Another result found in this paper concerns the unstable branches
mentioned above. When the gravitational coupling is turned on in the
Skyrme model without the sextic term -- corresponding to taking
gravitational backreaction into account -- then the unstable branches
of solutions smoothly converge back to the flat space Skyrmion
solution in the limit of vanishing black hole size.
Once we turn on the sextic term (sixth-order derivative term), there
is a small, but finite, critical value for the coefficient of said
term, for which the unstable branches end at a finite horizon radius:
hence the limit is not smooth. 

The paper is organized as follows.
Sec.~\ref{sec:model} introduces the model and the governing equations 
for the black hole in a Schwarzschild metric with Skyrme(-like) hair.
Sec.~\ref{sec:numerical} presents the numerical results and finally
Sec.~\ref{sec:discussion} concludes with a discussion. 

\section{The model}\label{sec:model}

The model is a nonlinear sigma model of Skyrme-type with
higher-derivative terms up to sixth order, coupled to gravity and the
action reads 
\begin{align}
S &= \int d^4x \; \sqrt{-g}\mathcal{L}, \qquad&
\mathcal{L} &= \mathcal{L}_S + \mathcal{L}_G, \\
\mathcal{L}_S &=
c_2\mathcal{L}_2
+ c_4\mathcal{L}_4
+ c_6 \mathcal{L}_6
- \delta c_0 V(U), \qquad&
\mathcal{L}_G &= \frac{1}{16\pi G} R,
\end{align}
where the $n$-th order Lagrangians are given by
\begin{align}
\mathcal{L}_2 &= -\frac{1}{4} g^{\mu\mu'} \tr(L_\mu L_{\mu'}),\\
\mathcal{L}_4 &= \frac{1}{32} g^{\mu\mu'} g^{\nu\nu'}
\tr\left(\big[L_\mu,L_\nu\big] \big[L_{\mu'},L_{\nu'}\big]\right),\\ 
\mathcal{L}_6 &= -\frac{1}{144} g_{\mu\mu'}(-g^{-1})
\big(\epsilon^{\mu\nu\rho\sigma}\tr\big[L_\nu L_\rho L_\sigma\big]\big)
\big(\epsilon^{\mu'\nu'\rho'\sigma'}\tr\big[L_{\nu'}L_{\rho'}L_{\sigma'}\big]\big),
\end{align}
where $L_\mu\equiv U^\dag\p_{\mu}U$ is the left-invariant current, 
$U=\sigma\mathbf{1}_2+i\pi^a\tau^a$, $a=1,2,3$ is the Skyrme field
with the constraint $\det U=1$, $g$ is the determinant of the metric
and we are using the mostly-negative signature of the metric.
$\mathcal{L}_2$ is the standard kinetic term, $\mathcal{L}_4$ is the
Skyrme term and $\mathcal{L}_6$ is the baryon current density squared,
which is inspired by the BPS Skyrme model \cite{Adam:2010fg}.

In the remainder of the paper, we will use the terminology
\begin{align}
  \mathrm{2+4\;model}:&\qquad c_4>0, \qquad c_6=0,\\
  \mathrm{2+4+6\;model}:&\qquad c_4>0, \qquad c_6>0.
\end{align}

The symmetry of $\mathcal{L}_S$ for $V=0$ is
$\tilde G =$ SU(2)$_{\rm L} \times $SU(2)$_{\rm R}$ acting on $U$ as 
$U \to U'= g_{\rm L} U g_{\rm R}^\dag$ and thus $L_\mu$ is manifestly
covariant. 
Finite energy configurations require that $U$ asymptotically takes on
a constant value, e.g.~$U=\mathbf{1}_2$.
Hence in the vacuum $\tilde{G}$ is spontaneously broken down to
$\tilde H \simeq$ SU(2)$_{\rm L+R}$, which in turn acts on $U$ as
$U \to U'= g U g^\dag$.
The target space is therefore
$\tilde G/\tilde H \simeq$ SU(2)$_{\rm L-R}$.

For concreteness we will use the potential
\beq
V(U) = \frac{1}{16}\tr\left[(2\mathbf{1}_2 - U - U^\dag)(2\mathbf{1}_2
  + U + U^\dag)\right],
\eeq
which is sometimes called the modified pion mass term
\cite{Kudryavtsev:1997nw,Weidig:1998ii,Piette:1997ce,Kudryavtsev:1999zm},
see also
Refs.~\cite{Nitta:2012wi,Nitta:2012rq,Gudnason:2013qba,Gudnason:2014nba,Gudnason:2014hsa,Gudnason:2014jga}. This 
potential breaks $\tilde{G}$ to SU(2)$_{\rm L+R}$ explicitly. 

The sixth-order term is written in a way where it is manifest that it
is the baryon current squared. It is however slightly easier to work
with the term after rewriting it in the following form
\cite{Gudnason:2014uha} 
\beq
\mathcal{L}_6 = \frac{1}{96}
g^{\mu\mu'}g^{\nu\nu'}g^{\rho\rho'}
\tr\big(L_\mu[L_\nu,L_\rho]\big)
\tr\big(L_{\mu'}[L_{\nu'},L_{\rho'}]\big).
\eeq

In this paper we will consider the Schwarzschild metric
\beq
ds^2 = N^2(r)C(r) dt^2
-\frac{1}{C(r)} dr^2
-r^2 d\theta^2 - r^2 \sin^2\theta d\phi^2,
\eeq
with
\beq
C(r) = 1 - \frac{2m(r)}{r},
\eeq
which is appropriate for studying a single Skyrmion -- which is also
spherically symmetric -- for which we choose the so-called hedgehog
Ansatz 
\beq
U = \cos f(r) + i\hat{x}^a\tau^a \sin f(r),
\eeq
where $\hat{x}$ is a spatial unit vector and $a=1,2,3$. 
Using the hedgehog Ansatz, we can write the static mass of the
Skyrmion as
\begin{align}
  M &= 4\pi\int_{r_h}^\infty dr \; r^2 N \bigg[
  c_2\left(\frac{1}{2} C f_r^2 + \frac{\sin^2 f}{r^2}\right)
  +c_4\frac{\sin^2 f}{r^2}\left(Cf_r^2 + \frac{\sin^2
    f}{2r^2}\right)
  + c_6 C\frac{\sin^4(f) f_r^2}{r^4} \non
&\phantom{=4\pi\int dr \; r^2 N \bigg[\ }
  +\frac{\delta c_0}{2}\sin^2 f
  \bigg],
\end{align}
where $f_r\equiv \p_r f$ and $r_h$ is the horizon radius. The mass of
the Skyrmion is the energy density integrated from the horizon to
infinity. 
We will now change variables to the dimensionless coordinate
$\rho\equiv\sqrt{\frac{c_0}{c_2}}r$ and rescale the coefficients
$c_4\to\frac{c_2^2}{c_0}c_4$ and $c_6\to\frac{c_2^3}{c_0^2}c_6$.
Although we have rescaled the coordinates by the coefficient of the
mass term, we insert a dimensionless mass, $\delta$ which can take the
value $0$ or $1$. If $\delta=0$ then $c_0$ is still the unit of the
would-be mass ($c_0$ can never vanish). In the case of $\delta=0$,
$c_0$ can be adjusted such that $c_4=1$. 
We can now write the mass as follows
\begin{align}
  M &= 4\pi\sqrt{\frac{c_2^3}{c_0}}
  \int_{\rho_h}^\infty d\rho \; \rho^2 N \bigg[
  \left(\frac{1}{2} C f_\rho^2 + \frac{\sin^2 f}{\rho^2}\right)
  +c_4\frac{\sin^2 f}{\rho^2}\left(Cf_\rho^2 + \frac{\sin^2
    f}{2\rho^2}\right)
  +c_6 C\frac{\sin^4(f) f_\rho^2}{\rho^4} \non
&\phantom{=4\pi\sqrt{\frac{c_2^3}{c_0}}\int_{\rho_h}^\infty d\rho \;
    \rho^2 N \bigg[\ }
  +\frac{\delta}{2}\sin^2 f
  \bigg],
\end{align}
where the dimensionless horizon radius is $\rho_h =
\sqrt{\frac{c_0}{c_2}}r_h$ and we define
\beq
\mu(\rho) = \sqrt{\frac{c_0}{c_2}} m(r).
\eeq
$c_4$, $c_6$ and $\delta=0,1$ are now dimensionless parameters. 

The baryon current is
\beq
\mathcal{B}^\mu =
-\frac{1}{24\pi^2}\frac{\epsilon^{\mu\nu\rho\sigma}}{\sqrt{-g}}
\tr(L_\nu L_\rho L_\sigma),
\eeq
and integrating the time component of this gives the baryon charge
\beq
B = \int d^3x\,\sqrt{-g}\; \mathcal{B}^0
= -\frac{2}{\pi}\int_{\rho_h}^\infty d\rho \; \sin^2(f) f_\rho
= \frac{2f(\rho_h) - \sin 2f(\rho_h)}{2\pi},
\eeq
and thus the total baryon charge $B$ is less than unity for any
$f(\rho_h)<\pi$ (we have used the asymptotic boundary condition
$f(\infty)=0$). 

The equation of motion for the Skyrme field profile $f$ is given by
\begin{align}
  &C \left(\rho^2 + 2c_4\sin^2f + \frac{2c_6\sin^4f}{\rho^2}\right)
  f_{\rho\rho}\non
&+\left[\left(C_\rho + C\frac{N_\rho}{N}\right)
  \left(\rho^2 + 2c_4\sin^2f + \frac{2c_6\sin^4}{\rho^2}f\right)
  +C\left(2\rho -\frac{4c_6 \sin^4f}{\rho^3}\right)\right] f_\rho\non
&+C\sin(2f) \left(c_4 + \frac{2c_6\sin^2f}{\rho^2}\right) f_\rho^2
-\sin(2f)\left(1 +\frac{c_4\sin^2f}{\rho^2} +
\frac{\delta\rho^2}{2}\right) = 0. \label{eq:f1}
\end{align}

The energy-momentum tensor can readily be calculated as
\begin{align}
T_{\mu\nu} &= -\frac{1}{2}\tr(L_\mu L_\nu)
  +\frac{c_4}{8}g^{\rho\sigma}\tr\left([L_\mu,L_\rho][L_\nu,L_\sigma]\right)
  +\frac{c_6}{16}g^{\rho\sigma}g^{\lambda\omega}
  \tr\left(L_\mu[L_\rho,L_\lambda]\right)\tr\left(L_\nu[L_\sigma,L_\omega]\right)
  \non
&\phantom{=\ }
  -g_{\mu\nu}\mathcal{L}_S.
\end{align}
Writing out the nonzero components, we have
\begin{align}
  \frac{1}{c_0}T_{tt} &= C N^2\left[
  \frac{1}{2}C f_\rho^2 + \frac{\sin^2 f}{\rho^2}
  +c_4\frac{\sin^2 f}{\rho^2}\left(C f_\rho^2 +
  \frac{\sin^2}{2\rho^2}\right)
  +c_6 C \frac{\sin^4(f) f_\rho^2}{\rho^4}
  +\frac{\delta}{2} \sin^2 f\right], \label{eq:Ttt}\\
  \frac{1}{c_0}T_{\rho\rho} &= 
  \frac{1}{2}f_\rho^2 - \frac{\sin^2f}{C\rho^2}
  +c_4\frac{\sin^2f}{\rho^2}\left(f_\rho^2-\frac{\sin^2f}{2C\rho^2}\right)
  +c_6\frac{\sin^4(f)f_\rho^2}{\rho^4}
  -\delta\frac{\sin^2f}{2C}, \label{eq:Trr}\\
  \frac{1}{c_0}T_{\theta\theta} &= \frac{1}{c_0}\frac{T_{\phi\phi}}{\sin^2\theta}
  = -\frac{1}{2} \rho^2 C f_\rho^2
  +c_4\frac{\sin^4f}{2\rho^2}
  +c_6 C\frac{\sin^4(f)f_\rho^2}{\rho^2}
  -\frac{\delta}{2}\rho^2\sin^2f. \label{eq:Tthetatheta}
\end{align}
We are now ready to obtain the Einstein equations
\beq
G_{\mu\nu} = 8\pi G T_{\mu\nu},
\eeq
and defining $\alpha\equiv 8\pi G c_0$, we can write down the
resulting equations by taking suitable linear combinations
\begin{align}
  \frac{1}{\alpha}\frac{N_\rho}{N} &= 
    \frac{1}{2}\rho f_\rho^2
    +c_4\frac{\sin^2(f)f_\rho^2}{\rho}
    +c_6\frac{\sin^4(f)f_\rho^2}{\rho^3}, \label{eq:N}\\
    \frac{1}{\alpha}C_\rho &= \frac{1-C}{\alpha\rho}
    -C\left(\frac{1}{2}\rho f_\rho^2
    +\frac{c_4\sin^2(f)f_\rho^2}{\rho}
    +\frac{c_6\sin^4(f)f_\rho^2}{\rho^3}\right)
    -\frac{\sin^2f}{\rho}
    -\frac{c_4\sin^4f}{2\rho^3}
    -\frac{\delta}{2}\rho\sin^2f. \label{eq:mu}
\end{align}
We can eliminate the field, $N$, by inserting Eq.~\eqref{eq:N}
into Eq.~\eqref{eq:f1} and simplify the coefficient of $f_\rho$ by
using Eq.~\eqref{eq:mu}. The resulting system of equations is then 
given by 
\begin{align}
  &C \left(\rho^2 + 2c_4\sin^2f + \frac{2c_6\sin^4f}{\rho^2}\right)
  f_{\rho\rho}\non
  &+\bigg[\left(1 - \alpha\sin^2f - \frac{\alpha c_4\sin^4f}{2\rho^2}
    - \frac{1}{2}\alpha\delta\rho^2\right)
    \left(\rho + \frac{2c_4\sin^2f}{\rho} +
    \frac{2c_6\sin^4f}{\rho^3}\right) \non
    &\qquad\qquad
    +C\left(\rho - \frac{2c_4\sin^2f}{\rho} -
    \frac{6c_6\sin^4f}{\rho^3}\right)\bigg] f_\rho \non
&+C\sin(2f) \left(c_4 + \frac{2c_6\sin^2f}{\rho^2}\right) f_\rho^2
-\sin(2f)\left(1 +\frac{c_4\sin^2f}{\rho^2} +
\frac{\delta\rho^2}{2}\right) = 0, \label{eq:f2}
\end{align}
and Eq.~\eqref{eq:mu}. 

In order to find numerical solutions to the system of equations
\eqref{eq:f2} and \eqref{eq:mu}, it will be convenient to use a
shooting method for ordinary differential equations (ODEs).
For that we need boundary conditions at the horizon (at $\rho_h$) with
a shooting parameter as well as boundary conditions at infinity.
The boundary condition at the horizon is
\beq
\lim_{\rho\to\rho_h} C = 1-\frac{2\mu(\rho_h)}{\rho_h} = 0,
\eeq
and hence $\mu(\rho_h) = \rho_h/2$.
$f(\rho_h)=f_h$ is the shooting parameter and by taking the limit
$\rho\to\rho_h$ of Eq.~\eqref{eq:f2}, we get 
\begin{align}
  C_\rho(\rho_h)
  \left(\rho_h^2 + 2c_4\sin^2f_h +
  \frac{2c_6\sin^4f_h}{\rho_h^2}\right)f_\rho(\rho_h)  
  -\sin(2f_h)\left(1 +\frac{c_4\sin^2f_h}{\rho_h^2}
  +\frac{\delta\rho_h^2}{2}\right) = 0. \label{eq:f1_rh}
\end{align}
Now we need to evaluate $C_\rho$ at the horizon
\begin{align}
  \lim_{\rho\to\rho_h}C_\rho =
  -\frac{2\mu_\rho(\rho_h)}{\rho_h} + \frac{1}{\rho_h}
  = -\alpha\left(\frac{\sin^2f_h}{\rho_h}
  +\frac{c_4\sin^4f_h}{2\rho_h^3}
  +\frac{\delta}{2}\rho_h\sin^2 f_h\right) +
  \frac{1}{\rho_h}, \label{eq:m1_rh} 
\end{align}
which follows straightforwardly from Eq.~\eqref{eq:mu}. 

Summarizing, we have the boundary conditions at the horizon
\begin{align}
  f(\rho_h) &= f_h, \\
  f_\rho(\rho_h) &= \frac{\rho_h^3\sin(2f_h)\left(2\rho_h^2 + 2c_4\sin^2f_h
    +\delta\rho_h^4\right)}{\left[
    2\rho_h^2-\alpha\sin^2 f_h\left(2\rho_h^2
  +c_4\sin^2f_h +\delta\rho_h^4\right)
  \right]\left(\rho_h^4 +
    2c_4\rho_h^2\sin^2f_h + 2c_6\sin^4f_h\right)},\\
  \mu(\rho_h) &= \frac{\rho_h}{2},
\end{align}
while at infinity they are
\beq
f(\infty) = 0, \qquad
\mu_\rho(\infty) = 0,
\eeq
where the second condition follows from the first and corresponds to
\beq
\lim_{\rho\to\infty}C = 1 - \frac{{\rm const}}{\rho},
\eeq
i.e., the metric is asymptotically Schwarzschild. 
In total there are exactly three boundary conditions on our
system (which is what is necessary) and one shooting parameter. 

Note that the first derivative of the Skyrmion profile function,
$f_\rho$, is negative at the horizon (as it should be), only when
\beq
\Xi \equiv 2\rho_h^2-\alpha\sin^2 f_h\left(2\rho_h^2
+c_4\sin^2f_h +\delta\rho_h^4\right),
\label{eq:Xidef}
\eeq
is positive, because $\sin 2f_h$ is negative for
$f_h\in(\tfrac{1}{2}\pi,\pi)$, which is the relevant range of the
shooting parameter.
When $\Xi$ vanishes, the first derivative is not defined on the
horizon and solutions cease to exist.
Notice that $\Xi=0$ corresponds to a vanishing Hawking temperature,
since
\beq
T_H = \frac{N(\rho_h)C_\rho(\rho_h)}{4\pi},
\eeq
and $\Xi=2\rho_h^3C_\rho(\rho_h)=0$ is equivalent to
$C_\rho(\rho_h)=0$ for $\rho_h>0$.

\section{Numerical solutions}\label{sec:numerical}

We will now pursue finding numerical solutions to the black hole
Skyrmion system. 
Eq.~\eqref{eq:f1_rh} implies that the coefficient of $f_{\rho\rho}$
vanishes at the horizon, which is problematic for a shooting
algorithm, as we would like to make a dynamic system of equations as
\beq
\p_\rho\begin{pmatrix}
f\\ f_\rho\\ \mu
\end{pmatrix} = M \begin{pmatrix}
  f\\ f_\rho\\ \mu
\end{pmatrix},
\eeq
where $M$ is some matrix (functional); the second row of the
right-hand side is defined by Eq.~\eqref{eq:f2} and the last row by
Eq.~\eqref{eq:mu}.  
However the right-hand side of the second row is not well defined at 
the horizon (the coefficient of $f_{\rho\rho}$ in Eq.~\eqref{eq:f2}
vanishes at the horizon).
Therefore we start the shooting from a very small radius
$\rho_\epsilon$ and calculate the values of the fields at
$\rho_h+\rho_\epsilon$ as
\begin{align}
  f(\rho_h+\rho_\epsilon) &= f_h + \rho_\epsilon f_\rho(\rho_h),\\
  f_\rho(\rho_h+\rho_\epsilon) &= f_\rho(\rho_h),\\
  \mu(\rho_h+\rho_\epsilon) &= \frac{\rho_h}{2} +
  \rho_\epsilon\mu_\rho(\rho_h),
\end{align}
where $\mu_\rho(\rho_h)$ is given by Eq.~\eqref{eq:m1_rh}.
This is a good approximation if $\rho_\epsilon$ is extremely small.
In the numerical calculations we have found that
$\rho_\epsilon\lesssim 10^{-5}$ is small enough for allowing for the
linear approximation and large enough to start the shooting
algorithm.

From $\rho_h+\rho_\epsilon$ we employ a standard fourth-order
Runge-Kutta method to integrate the equations
(\ref{eq:mu}-\ref{eq:f2}) up to an appropriately chosen cutoff. 

We are now ready to present the numerical results.
We start by reproducing the well-known results in the 2+4 model, which
is simply the standard Skyrme model with the addition of the modified
pion mass. 
After rescaling we have two free parameters: the gravitational
coupling $\alpha$ and $c_4$. Actually, if we were to consider the
model without the potential term, then the rescaling would eliminate
$c_4$ instead of the potential parameter (pion mass)\footnote{Of
  course this mass parameter is not related to the mass of the
  physical pion in QCD. }.

\begin{figure}[!htp]
  \begin{center}
    \mbox{\subfloat[]{\includegraphics[width=0.49\linewidth]{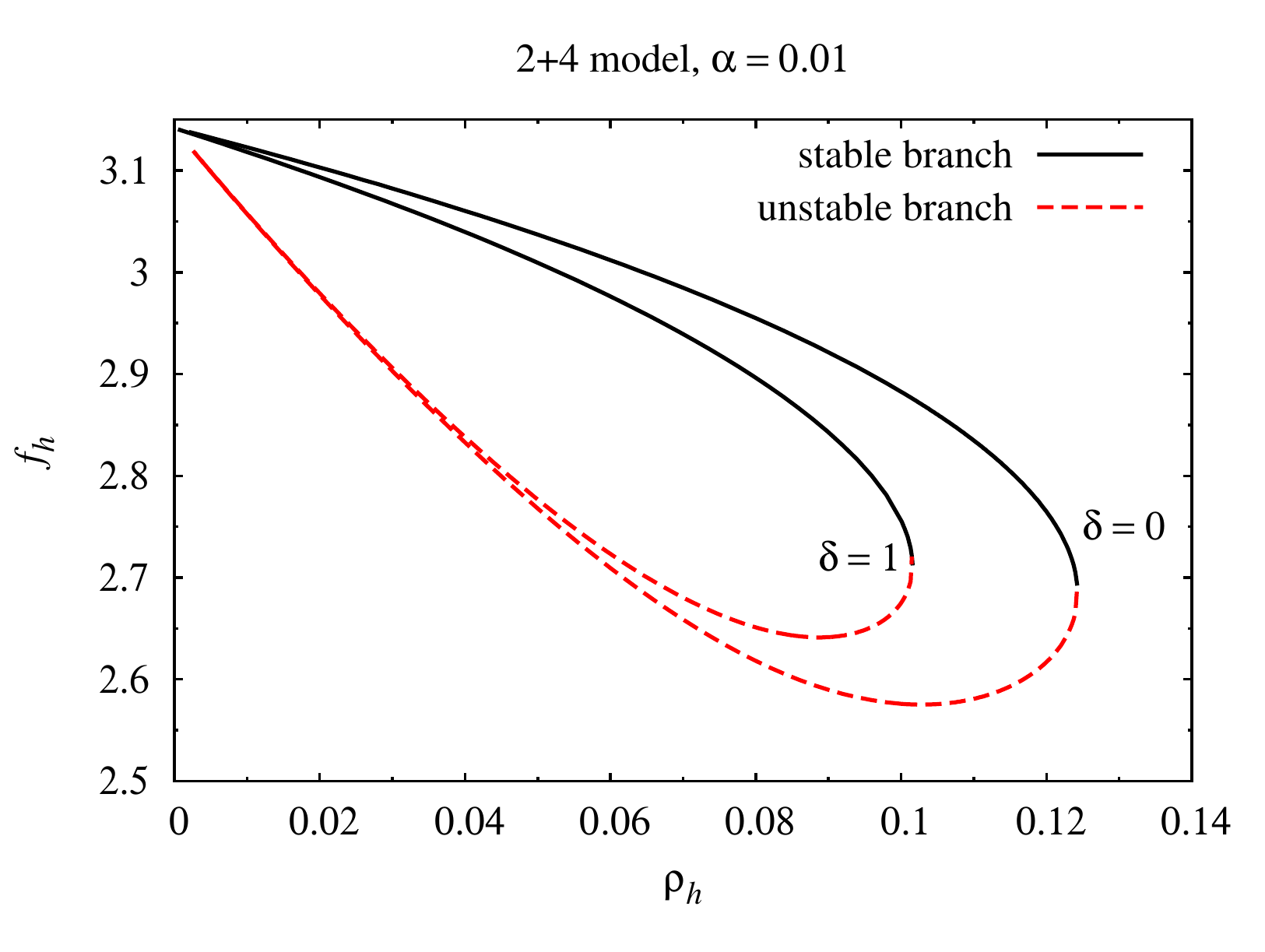}}
      \subfloat[]{\includegraphics[width=0.49\linewidth]{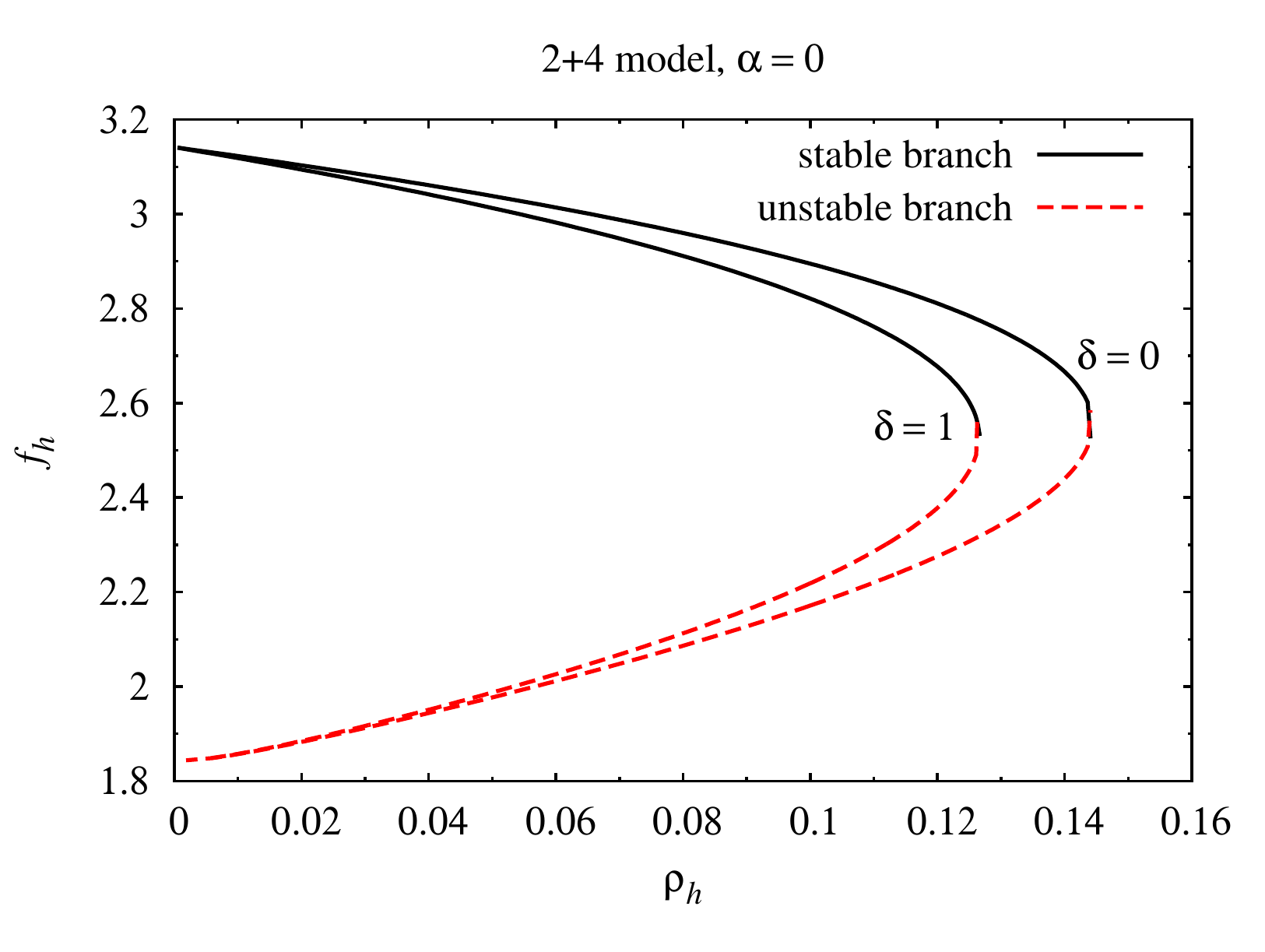}}}
    \caption{Branches of solutions for the 2+4 model with and without
      mass term ($\delta=0,1$) for (a) gravitational coupling
      $\alpha=0.01$ and (b) $\alpha=0$. In this figure $c_4=1$. }
    \label{fig:m24}
  \end{center}
\end{figure}

In Fig.~\ref{fig:m24} we show the stable and unstable branches in
black solid lines and dashed red lines, respectively, and for all
combinations of vanishing/nonvanishing gravitational coupling and
vanishing/nonvanishing potential parameter $\delta$ (recall that
$\delta$ after rescaling, can only take the values 0 or 1). 
The first thing we note is, as explained in the introduction, that
with the gravitational coupling turned on, the unstable branch
smoothly converges back towards the flat space Skyrmion solution as
the horizon radius $\rho_h$ is sent to zero; whereas in the case of
vanishing gravitational coupling it does not. The unstable branch
continues down in the $f_h$ direction and in the limit of vanishing 
$\rho_h$ the solution is discontinuous and ceases to exist
\cite{Luckock:1986tr}.

\begin{figure}[!htp]
  \begin{center}
    \mbox{\subfloat[]{\includegraphics[width=0.49\linewidth]{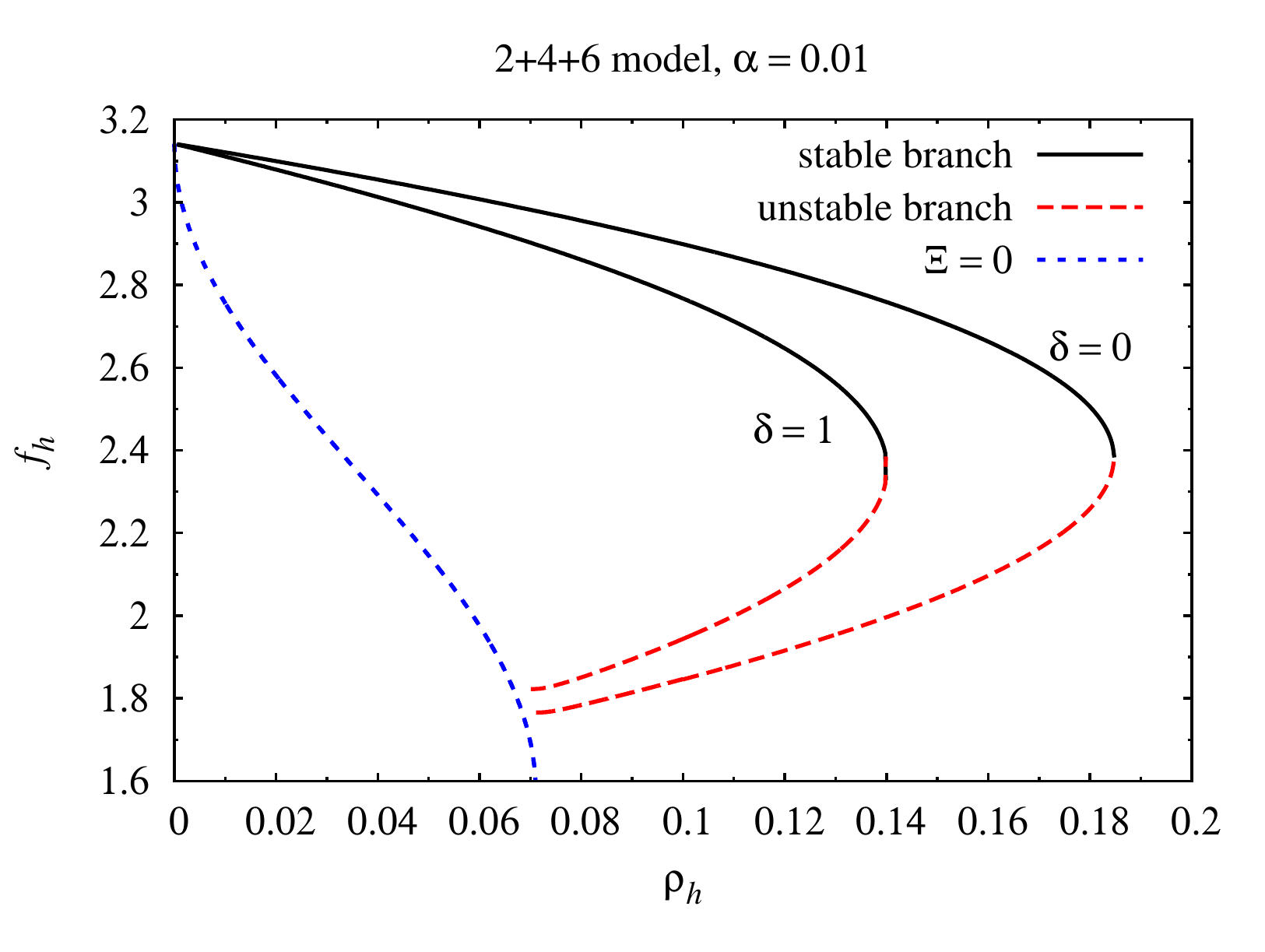}}
    \subfloat[]{\includegraphics[width=0.49\linewidth]{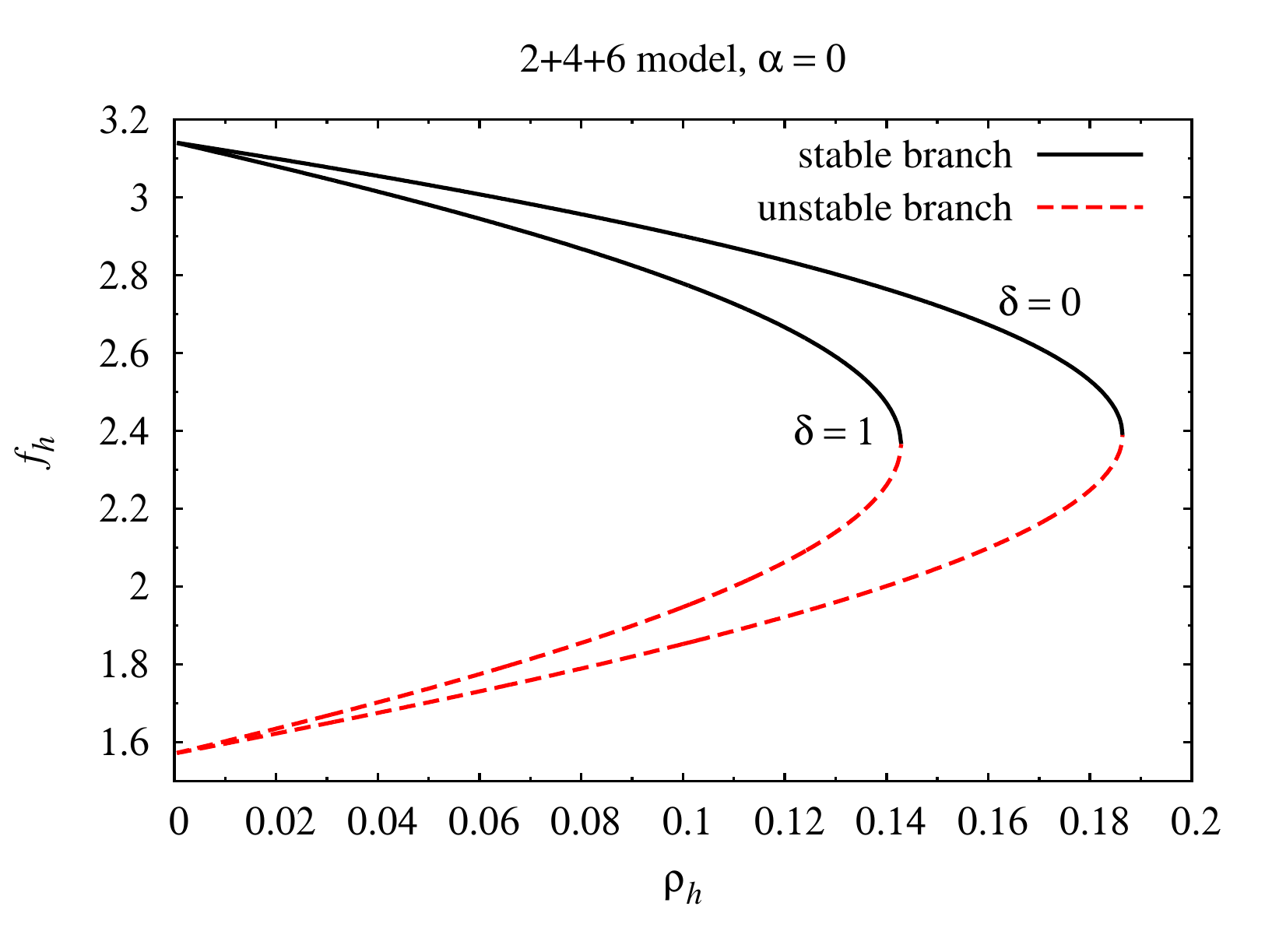}}}
    \caption{Branches of solutions for the 2+4+6 model with and without
      mass term ($\delta=0,1$) for (a) gravitational coupling
      $\alpha=0.01$ and (b) $\alpha=0$.
      The blue dashed line to the left in figure (a) represents the
      vanishing of $\Xi=0$ of Eq.~\eqref{eq:Xidef} which corresponds
      to a vanishing Hawking temperature, at which there is no black
      hole hair solution. 
      In this figure $c_4=c_6=1$. }
    \label{fig:m246}
  \end{center}
\end{figure}

We now turn on a positive value for the coefficient of the sextic
term, $c_6>0$; this corresponds to the 2+4+6 model.
\emph{A priori} one would not expect substantial differences with
respect to the 2+4 model discussed above, because in flat space the
soliton solutions (in this parameter range; that is, when $c_4$ is of
the same order of magnitude as $c_6$ or larger) are quite similar
\cite{Gudnason:2015nxa,Gudnason:2014jga,Gudnason:2014hsa}. 
However, by inspection of Fig.~\ref{fig:m246} we see that differences
in the unstable branches, with respect to the 2+4 model, emerged.
The stable branches are quite similar to that of the 2+4 model, except
that they are longer; i.e.~the Skyrme hair solutions in the 2+4+6
model are stable for much larger black holes than the 2+4 model. 
The unstable branches without the gravitational coupling turned on
remain similar to those of the 2+4 model; however, when the
gravitational backreaction is turned on, they do not converge back
towards the flat space Skyrmion solution.
In some sense the unstable branches are on the same trajectory
(downwards in the $(\rho_h,f_h)$ phase diagram) as those without
gravitational backreaction for decreasing horizon radius, $\rho_h$.
However, long before reaching the limit of vanishing black hole size,
the unstable branches come close to the $\Xi=0$ line in the diagram,
where $\Xi$ is defined in Eq.~\eqref{eq:Xidef}.
This phenomenon happens both with and without the potential term.
The $\Xi=0$ line is defined by the position of the pole in the first
radial derivative of the Skyrmion profile function $f_\rho(\rho_h)$
at the horizon radius, and corresponds to a vanishing Hawking
temperature.
It is intuitively clear that when the Skyrmion profile blows up at the
black hole horizon, no continuous stable black hole hair solution
exists.
It is also clear that if the Hawking temperature vanishes for
a finite black hole mass, then the entropy would have to blow up; this
should not happen in a physical system and hence it signals an
instability of the black hole hair.
We observe from Fig.~\ref{fig:m246} that the unstable branch of
solutions ceases to exist slightly before $\Xi=0$, but quite close to
this line in the diagram. 

\begin{figure}[!htp]
  \begin{center}
    \includegraphics[width=0.49\linewidth]{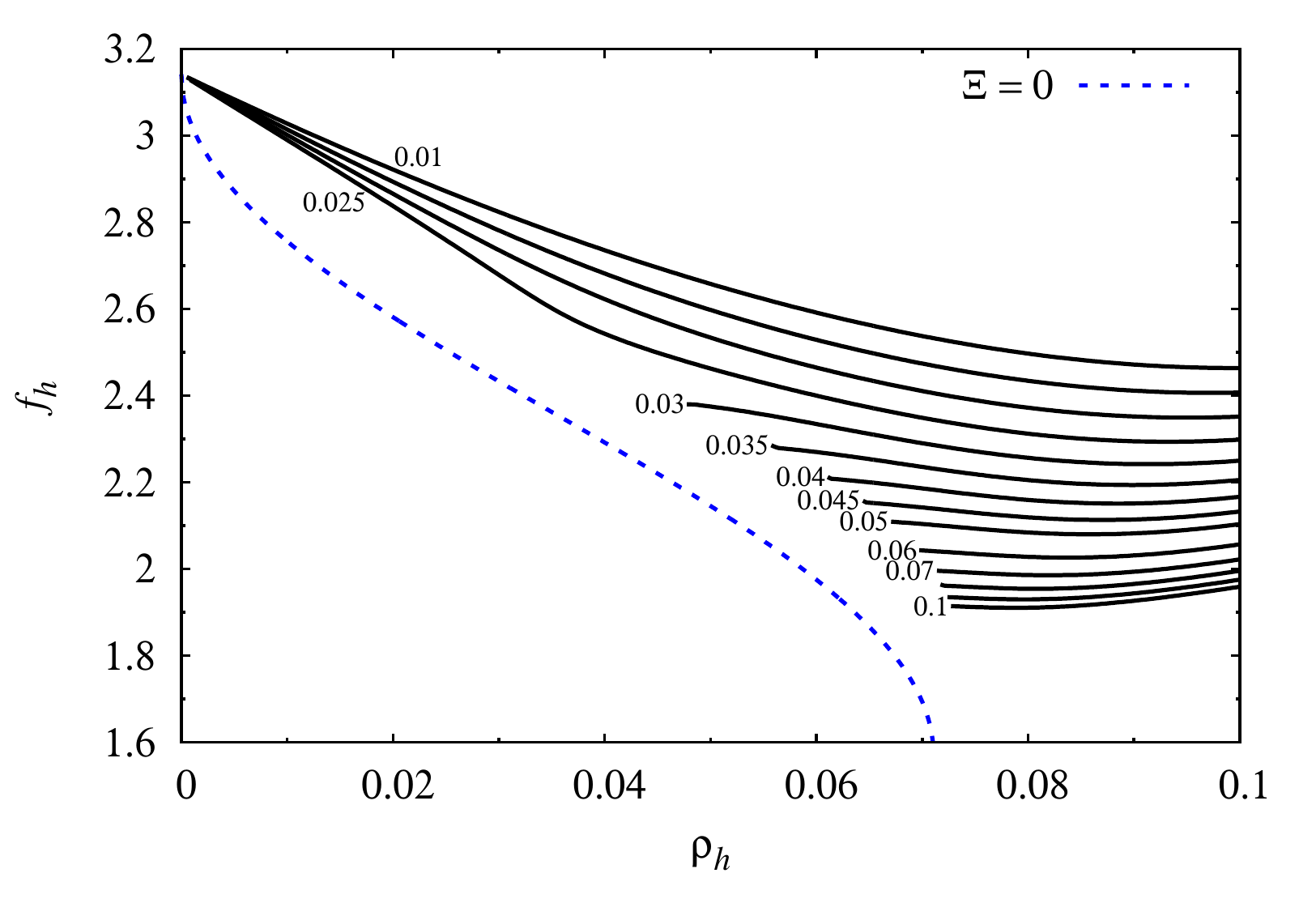}
    \caption{Unstable branches of solutions for the 2+4+6 model
      without mass term for various values of
      $c_6=0.01,0.015,0.02,0.025,0.03,0.035,0.04,0.045,0.05,0.06,0.07,0.08,0.09,0.1$. The
      values of $c_6$ are indicated on the figure.
      The blue dashed line shows where $\Xi=0$. 
      In this figure $c_4=1$, $\delta=0$ and the gravitational 
      coupling is $\alpha=0.01$.} 
    \label{fig:m246_unstable}
  \end{center}
\end{figure}

As we know that in the 2+4 model, the unstable branch moves upwards in
the $(\rho_h,f_h)$ phase diagram for decreasing horizon radius
$\rho_h$, there should be some critical value of $c_6$ for which the
unstable branches start to end at a finite horizon radius $\rho_h>0$.
We therefore consider taking the limit of $c_6\to 0$ and see when the
unstable branches start to exist in the limit of $\rho_h\to 0$. 
Fig.~\ref{fig:m246_unstable} shows the phase diagram with only the
unstable branches for various values of the coefficient of the sextic
term, $c_6$, and we can see from the figure that for $c_4=1,\delta=0$,
the critical value of $c_6$ is between $0.025$ and $0.03$. 

\begin{figure}[!htp]
  \begin{center}
    \includegraphics[width=0.49\linewidth]{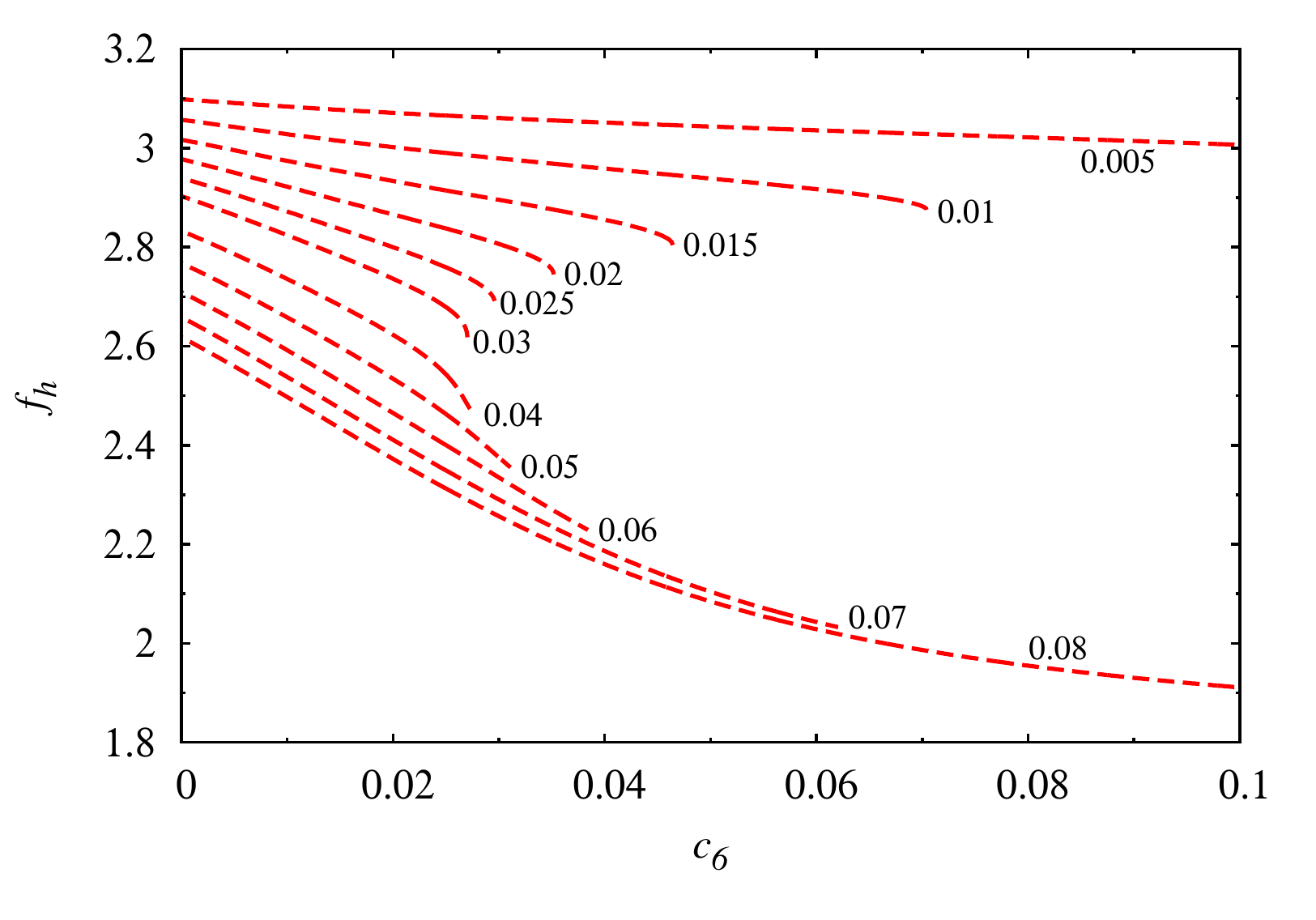}
    \caption{Families of unstable solutions for the 2+4+6 model for various
      horizon radii,
      $\rho_h=0.005,0.01,0.015,0.02,0.025,0.03,0.04,0.05,0.06,0.07,0.08$ as
      functions of $c_6$. The horizon radii are indicated on the
      figure. 
      In this figure $c_4=1$, $\delta=0$ and the gravitational
      coupling is $\alpha=0.01$. } 
    \label{fig:m246_c6branch}
  \end{center}
\end{figure}

In Fig.~\ref{fig:m246_c6branch} we show the same physics, but in terms
of $c_6$ and $f_h$; the different curves depict various horizon radii,
$\rho_h$. This figure clearly shows that if we turn off the sextic
term, $c_6=0$, then all (the shown) horizon radii have solutions.
Curiously, the lines open up in the limit of $\rho_h\to 0$ and allow
for bigger $c_6$ than for example $\rho_h=0.03$, which is the most
restricting radius in the diagram.
From Fig.~\ref{fig:m246_unstable} we estimated that the critical value
of $c_6$ for which the unstable branch ends at a finite horizon
radius, $\rho_h>0$, is about $0.025$-$0.03$; while from
Fig.~\ref{fig:m246_c6branch}, we can confirm that it is slightly less
than $0.03$, so in accord with the previous estimate. 

\begin{figure}[!htp]
  \begin{center}
    \mbox{\subfloat[$c_6=1$]{\includegraphics[width=0.49\linewidth]{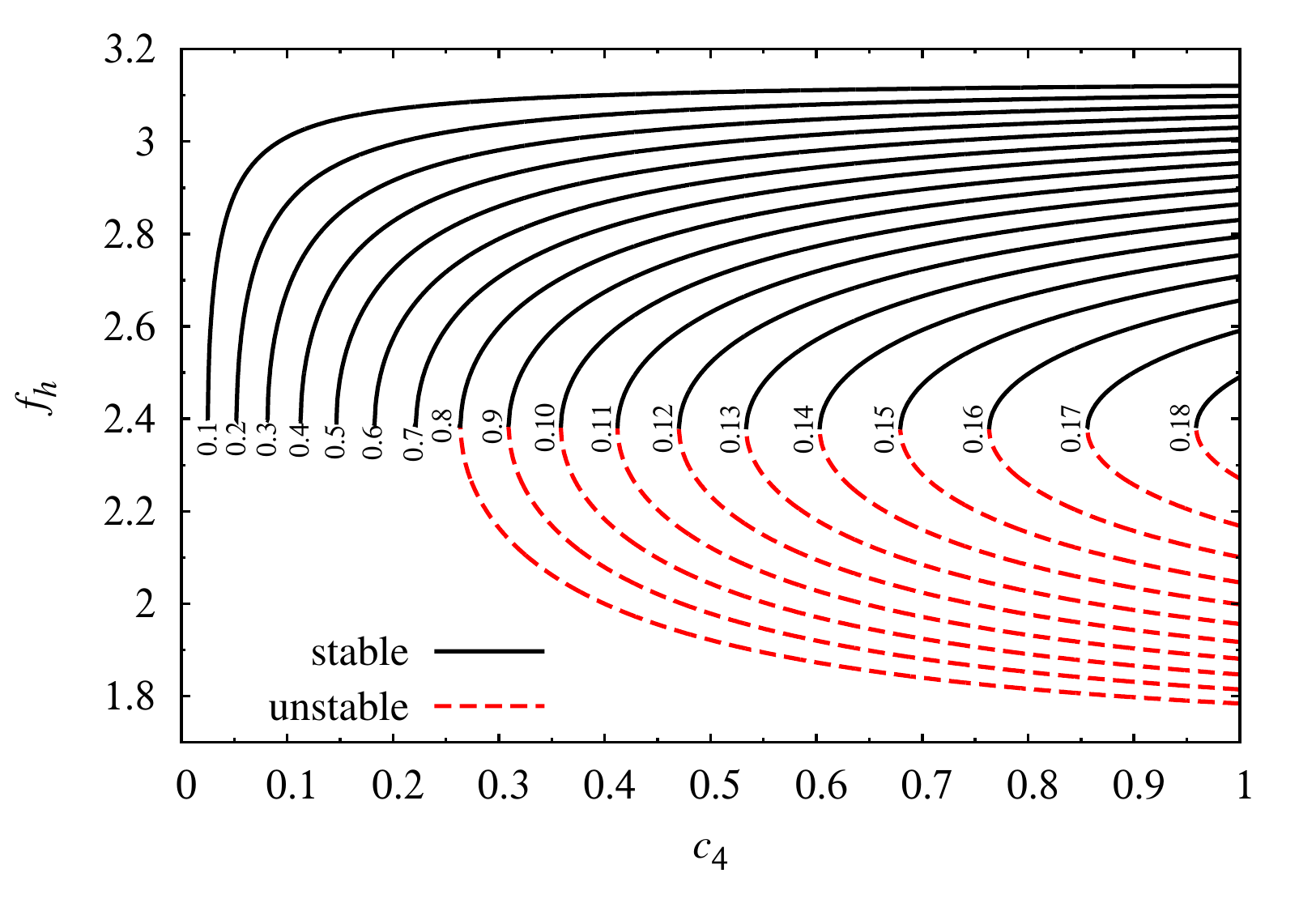}}
    \subfloat[$c_6=0$]{\includegraphics[width=0.49\linewidth]{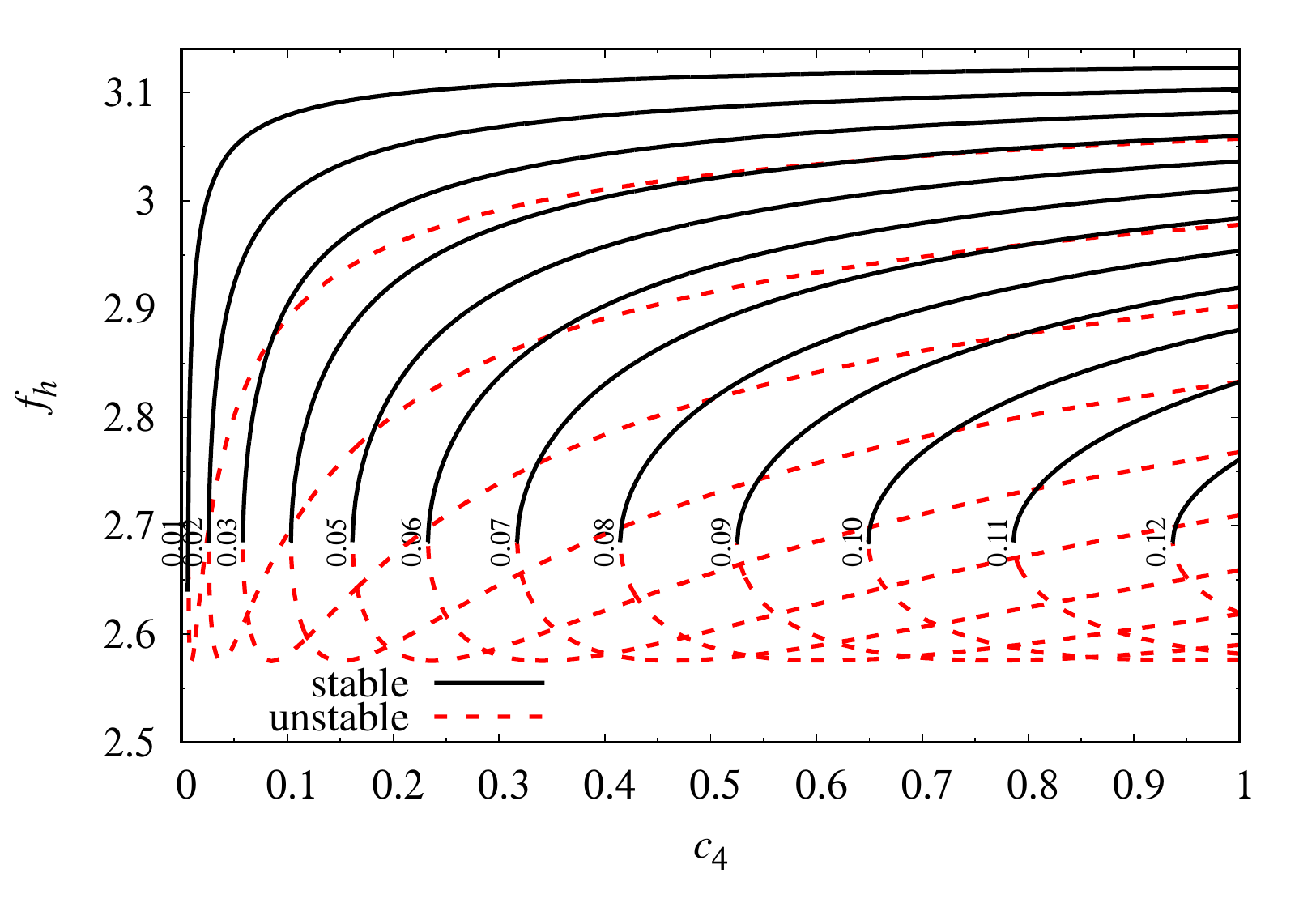}}}
    \caption{Families of solutions for (a) the 2+4+6 model and (b) the
    2+4 model, for various horizon radii,
      (a)
    $\rho_h=0.01$, $0.02$, $0.03$, $0.04$, $0.05$, $0.06$, $0.07$,
    $0.08$, $0.09$, $0.10$, $0.11$, $0.12$, $0.13$, $0.14$, $0.15$,
    $0.16$, $0.17$, $0.18$
    and (b)
    $\rho_h=0.01,0.02,0.03,0.04,0.05,0.06,0.07,0.08,0.09,0.10,0.11,0.12$
      as functions of $c_4$. The horizon radii are indicated on the
      figure. We can see from the figure that for decreasing  
      values of $c_4$, only smaller and smaller horizon radii exist --
      even on the stable branch.
      In this figure $\delta=0$ and the gravitational
      coupling is $\alpha=0.01$. 
      } 
    \label{fig:m246_c4branch_stable}
  \end{center}
\end{figure}

Our final numerical investigation considers turning off the Skyrme
term for fixed coefficient of the sextic term $c_6=1$.
Fig.~\ref{fig:m246_c4branch_stable}a shows both stable and unstable
solutions in the ($c_4,f_h$) diagram for various horizon radii
$\rho_h$.
For comparison, we show also the analogous figure $(c_4,f_h)$
for $c_6=0$ in Fig.~\ref{fig:m246_c4branch_stable}b where it is clear
that the black hole Skyrme hair will cease to exist when the Skyrme
term is turned off. The biggest difference between the two panels lies
in the unstable branches, because in the 2+4 model the unstable
branches return to the flat-space Skyrmion when the black hole horizon
radius is sent to zero ($\rho_h\to 0$).
Fixing $c_4$ we can read off the $f_h$ branch as function of the
horizon radii by looking at a vertical line in the figure.
Following a fixed horizon radius, $\rho_h$, we can see at which value
of the Skyrme term coefficient, $c_4$, the solutions cease to exist.
Interestingly -- and this is one the main findings of this paper --
\emph{all} solutions cease to exists in the limit of $c_4\to 0$, even
though we have the sextic term turned on. 
We can also see that the unstable branches cease to exist quite before
the stable branches (in the case of $c_6=1$, see
Fig.~\ref{fig:m246_c4branch_stable}a).
We can physically understand that the bifurcation point -- which is
the maximal size of black hole that can support the Skyrme hair --
simply goes to zero in the limit of $c_4\to 0$ for fixed $c_6=1$; we 
can perhaps say that the black hole eats the Skyrme hair if there is
no Skyrme term turned on.

\section{Discussion and conclusion}\label{sec:discussion}

In this paper we have considered Schwarzschild black holes with Skyrme
hair in a Skyrme-like model with the addition of a sixth-order
derivative term as well as a potential term.
We first reproduce the expected branches of solutions in the
$(\rho_h,f_h)$ phase diagram for the Skyrme model with the sextic term
turned off.
Then turning on the sextic term, we find that the unstable branches
are modified and end at finite horizon radii -- beyond which no
unstable solution exists, for all but very small values of the
coefficient of the sextic term.
This is due to the unstable branches coming too close to a
line in the phase diagram ($\Xi=0$) where the Hawking temperature
vanishes.
Furthermore, we find the quite surprising result that there is no
stable or unstable Skyrme hair for any black holes -- with or without
the potential -- in the limit of vanishing coefficient of the Skyrme
term, $c_4\to 0$.
This unexpected result implies that, although higher-derivative terms
can stabilize Skyrmions in flat space, the sixth-order derivative
term cannot stabilize the Schwarzschild black hole hair. 

In Ref.~\cite{Gudnason:2015dca} we observed that there are no black 
holes in the BPS-Skyrme submodel; i.e.~in the case without a kinetic
term and without the Skyrme term. This observation was made in the
case of a particular potential and so it was not clear whether
Skyrme hair solutions in the 2+6 model would be stable or not.
Now we have the answer in the negatory. 
This interesting result begs for the question: under what
circumstances does black hole hair exist?
It is quite unlikely that any potential of any type will alter this
conclusion as potentials tend to collapse the Skyrmions and the black
hole already does that without the Skyrme term present -- even with
the sixth-order derivative term.
Let us note that for small values of the Skyrme term
coefficient (small $c_4$), the unstable branches come too close to the
line in the phase diagram where the Hawking temperature vanishes.
Furthermore, we observe that the critical point moves to smaller and
smaller black hole horizons for decreasing values of $c_4$ and
eventually leads to a vanishing black hole size even for the stable
branches. This in turn means that the \emph{stable branch} is
approaching the line where the Hawking temperature is vanishing.
Let us also note that the sixth-order derivative term itself --
without the presence of the second-order kinetic term -- leads to a
theory described by a perfect fluid \cite{Adam:2014nba}.
Combining these two facts, we can intuitively understand that the
black hole Skyrme hair with a sixth-order derivative term --
possessing the properties of a perfect fluid -- cannot withstand the
gravitational attraction: the black hole Skyrme hair collapses.

It is not clear at this point if all higher-order derivative terms
higher than fourth order will be unable to stabilize black hole Skyrme
hair.
Higher-order terms may not yield a theory with the properties
of a perfect fluid.
We will leave this question for future studies.
We conjecture that the sixth-order derivative term, due to its
properties of a perfect fluid, is the only higher-order derivative
term leading to a second-order radial equation of motion which cannot
stabilize the black hole Skyrme hair.
The proof thereof awaits to be found.

\subsection*{Note added}
While this manuscript was under preparation we were
informed\footnote{We thank Andrzej Wereszczynski for letting us
  know. } that a paper with similar results was about to appear on
the arXiv \cite{Adam:2016vzf}.

\subsection*{Acknowledgments}

S.~B.~G.~thanks the Recruitment Program of High-end Foreign
Experts for support.
The work of M.~N.~is supported in part by a Grant-in-Aid for
Scientific Research on Innovative Areas ``Topological Materials
Science'' (KAKENHI Grant No.~15H05855) and ``Nuclear Matter in Neutron
Stars Investigated by Experiments and Astronomical Observations''
(KAKENHI Grant No.~15H00841) from the the Ministry of Education,
Culture, Sports, Science (MEXT) of Japan. The work of M.~N.~is also
supported in part by the Japan Society for the Promotion of Science
(JSPS) Grant-in-Aid for Scientific Research (KAKENHI Grant
No.~16H03984) and by the MEXT-Supported Program for the Strategic
Research Foundation at Private Universities ``Topological Science''
(Grant No.~S1511006).

\end{document}